\documentclass[twocolumn,superscriptaddress,amsmath,ammsymb]{revtex4-1}
\usepackage{epsfig}
\usepackage{amsmath}
\usepackage{amssymb}
\usepackage{graphicx}
\usepackage{dcolumn}
\usepackage{bm}
\usepackage[usenames]{color}
\usepackage[breaklinks]{hyperref}
\usepackage{soul}
\usepackage{ulem}
\hypersetup{colorlinks=true,allcolors=blue}

\begin{document}
	
\title{Unconventional optical selection rules in ZrTe$_5$ under an in-plane magnetic field}

\author{Yi-Xiang Wang}
\email{wangyixiang@jiangnan.edu.cn}
\affiliation{School of Science, Jiangnan University, Wuxi 214122, China}
\affiliation{School of Physics and Electronics, Hunan University, Changsha 410082, China}

\author{Fuxiang Li}
\email{fuxiangli@hnu.edu.cn}
\affiliation{School of Physics and Electronics, Hunan University, Changsha 410082, China}
 
\date{\today}

\begin{abstract} 
The optical selection rules of an electron system under a magnetic field play key roles in determining its optical properties, from which the band structures and underlying symmetries can be derived.  In this paper, based on a three-dimensional strong topological insulator model describing ZrTe$_5$, we study the Landau levels (LLs) and magneto-optical conductivity under an in-plane magnetic field.  We reveal that in the transverse conductivity Re$(\sigma_{zz})$, the unconventional optical selection rules $n\rightarrow n\pm2$ dominate, with $n$ being the LL index.  We  attribute the unconventional selection rules to the peculiar distribution of parity carried by the LLs, resulting from the chiral symmetry of the sub-Hamiltonians.  Moreover, we predict that, if the strong anisotropic system is tuned to be nearly isotropic, the LLs would redistribute and the conventional selection rules $n\rightarrow n\pm1$ can be recovered.   
\end{abstract} 

\maketitle

\section{Introduction} 
Three-dimensional (3D) zirconium pentatelluride ZrTe$_5$ is a representative of emerging topological materials and has aroused researchers' ongoing interests in recent years~\cite{Lv}.  Its nontrivial band topology~\cite{Weng} together with the intrinsic dynamics of the Dirac fermion can bring out a lot of unconventional phenomena, such as the chiral magnetic effect~\cite{Q.Li}, the anomalous Hall effect~\cite{T.Liang, Y.Liu}, the 3D quantum Hall effect~\cite{F.Tang}, and the anomalous thermoelectric effect~\cite{J.L.Zhang2019, W.Zhang}.  Although there is no consensus on the ground state of ZrTe$_5$ in experiment, recent measurements of angle-resolved photoelectron spectroscopy~\cite{Manzoni}, infrared spectroscopy~\cite{Z.G.Chen, B.Xu}, magnetoinfrared spectroscopy~\cite{Y.Jiang2020} as well as magneto-transport~\cite{J.Wang} supported its ground state to be a strong topological insulator (TI) at low temperatures, in which the characteristic linear surface states~\cite{Manzoni} and the bulk band inversions have been clearly identified~\cite{Z.G.Chen, B.Xu, Y.Jiang2020, J.Wang}. 

In 3D topological Weyl/Dirac semimetal materials, applying a strong magnetic field can lead to the formation of Landau levels (LLs) dispersing along the wave vector parallel to the field.  When combining the magnetic field with scanning tunneling spectroscopy or infrared spectroscopy, they can provide powerful ways to study the bulk electronic structure and band topology~\cite{Jeon, X.B.Li}.  Among them, the optical selection rules play key roles in determining the basic optical properties of the electron system~\cite{Akrap, R.Y.Chen, E.Martino}.  In a 3D ZrTe$_5$ crystal, prismatic ZrTe$_3$ chains run along the crystallographic $a$ axis ($x$ axis) and link along the $c$ axis ($y$ axis) with the zigzag Te atom chains, which form two-dimensional (2D) layers.  Via weak van-der Waals bondings, the 2D layers stack along the $b$ axis ($z$ axis) to form a 3D layered orthorhombic structure. 
In a recent magnetoinfrared spectroscopy experiment of ZrTe$_5$~\cite{Y.Jiang2020}, the magnetic field was applied not only along the perpendicular $z$ direction, but also in the $x$-$y$ plane.  Since the LLs and magneto-optics under an in-plane magnetic field remain unexplored in quantum theory, this motivates the present work. 

In this paper, based on a strong TI model describing ZrTe$_5$, we study the LLs and magneto-optical conductivity Re$(\sigma_{\alpha\alpha})$ under an in-plane magnetic field.  By performing numerical calculations in the lattice model, we find that the zeroth LLs show distinctive behaviors from the $n\geq1$ LLs, leading to different resonant peak numbers in Re$(\sigma_{\alpha\alpha})$, which are consistent with  experimental observations~\cite{Y.Jiang2020}.  In the transverse conductivity Re$(\sigma_{zz})$, we reveal the existence of unconventional optical selection rules $n\rightarrow n\pm2$ and attribute them to the peculiar distribution of parity carried by the LLs.  Such a distribution is shown to be induced by  the chiral symmetry of  the sub-Hamiltonians with quadratic terms.  Moreover, we predict that, if a strong anisotropic system is tuned to be nearly isotropic under an external pressure, the LLs will redistribute and the conventional selection rules $n\rightarrow n\pm1$ can be recovered.  Our findings will broaden the understanding of exotic quantum phenomena in topological materials.

\section{Model}
We start from the strong TI model of ZrTe$_5$ that was proposed in a recent magnetoinfrared spectroscopy experiment~\cite{Y.Jiang2020}.  The corresponding low-energy Hamiltonian can be written as $(\hbar=1)$~\cite{Y.Jiang2020, H.Zhang, C.X.Liu, Wang2021, L.You}  
\begin{align}
H(\boldsymbol k)
&=v(k_x\sigma_z\otimes\tau_x+k_yI\otimes \tau_y)
+v_z k_z\sigma_x\otimes\tau_x
\nonumber\\
&+[M-\xi(k_x^2+k_y^2)-\xi_zk_z^2]I\otimes \tau_z, 
\label{Hk}
\end{align}
where $\sigma$ and $\tau$ are the Pauli matrices acting on the spin and orbit degrees of freedom, respectively.  $v$ and $v_z$ are the Fermi velocities, $\xi$ and $\xi_z$ are the band inversion parameters, and $M$ denotes the Dirac mass.  Consider an in-plane magnetic field, $\boldsymbol B=B\boldsymbol e_x$.  To incorporate it in the system, we use the Peierls substitution, $\boldsymbol k\rightarrow \boldsymbol k-e\boldsymbol A$, where the magnetic vector potential is chosen as $\boldsymbol A=By\boldsymbol e_z$ in the Landau gauge.  In the following calculations, unless specified, we take the model parameters as in Ref.~\cite{Y.Jiang2020}: $(v,v_z)=(6,0.5)\times10^5$ m/s, $(\xi,\xi_z)=(100,200)$ meV nm$^2$, and $M=7.5$ meV.

\section{Main Results} 

\subsection{LLs}

\begin{figure}
	\includegraphics[width=9.2cm]{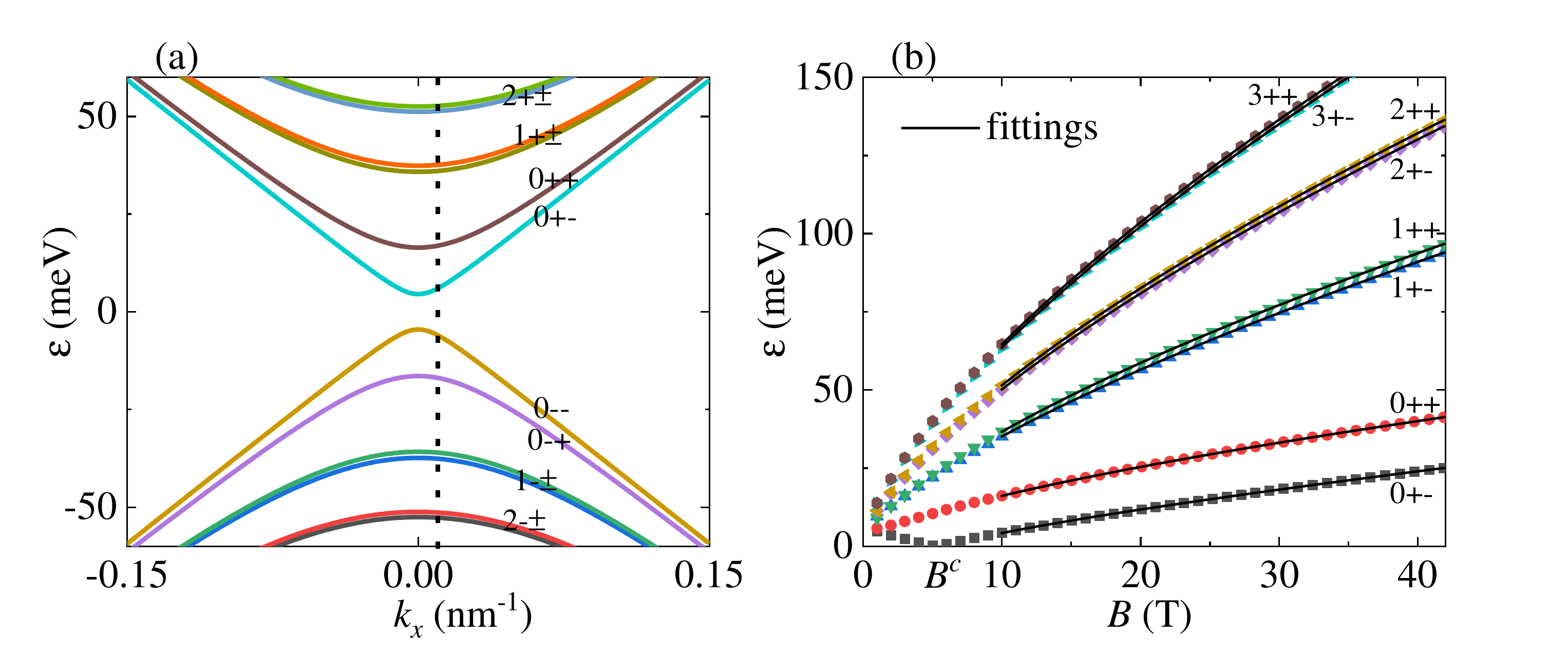}
	\caption{(Color online) The LL dispersions in (a), with  magnetic field $B=10.3$ T.  The LL energy $\varepsilon_{ns\lambda}$ at $k_x=0$ vs $B$ in (b).  The solid lines are the fittings to the LLs by the power function $\varepsilon=(aB+c)^d$ for $B>10$ T.  From $(0+-)$ to $(3++)$ LL, the exponent $d$ is extracted as $d$=0.80, 0.66, 0.69, 0.68, 0.68, 0.68, 0.68, 0.68. }
	\label{Fig1}	
\end{figure}

To solve the LLs, we construct the tight-binding Hamiltonian in the lattice model [see Sec. I of the Supplemental Material (SM) \cite{SM}]. 
By diagonalizing the magnetic unit cell, the LL dispersions can be obtained and are shown in Fig.~\ref{Fig1}.  The index $ns\lambda$ are used to label each LL, with $s=\pm1$ denoting the conduction/valence band and $\lambda=\pm1$ the two branches.
  
Figure~\ref{Fig1}(a) shows that for the zeroth LLs, the two branches are well separated in energy, while for the $n\geq1$ LLs, the two branches are almost degenerate.  At strong magnetic field, all LLs increase with a relation that can be fitted by the power formula $\varepsilon=(aB+c)^d$ as shown in Fig.~\ref{Fig1}(b).  For the lowest LL, the exponent $d=0.80$, whereas for other LLs, the exponents are close to $\frac{2}{3}$, both of which are distinct from the characteristic exponent $d=\frac{1}{2}$ of the LLs in a 3D linear Dirac model~\cite{Armitage}. 
We note that the lowest $(0+-)$ LL shows anomalous behaviors with the magnetic field: It first decreases, reducing to zero at a critical $B^c$, and then increases.  This anomalous behavior will be analyzed later.  
  
\subsection{Magneto-optical Conductivity}

The LL structure can be probed by magneto-optical measurements.  Within the linear-response theory, the magneto-optical conductivity $\sigma_{\alpha\alpha}$ is calculated by using the Kubo formula~\cite{M.Udagawa}, 
\begin{align}
\sigma_{\alpha\alpha}(\omega)=&\frac{1}{iV}
\sum_{nn'}\sum_{ss'}\sum_{\lambda\lambda'} 
\frac{f(\varepsilon_{ns\lambda})-f(\varepsilon_{n's'\lambda'})}
{\varepsilon_{ns\lambda}-\varepsilon_{n's'\lambda'}} 
\nonumber
\\
&
\times\frac{|\langle\psi_{ns\lambda}|J_\alpha|\psi_{n's'\lambda'}\rangle|^2}
{\omega+\varepsilon_{ns\lambda}-\varepsilon_{n's'\lambda'}+i\tau^{-1}}, 
\label{Kubo}
\end{align}
where $V$ is the volume of the system, $f(x)$ is the Fermi-Dirac distribution function, $J_\alpha=-ie[r_\alpha,H]$ is the current density operator, $\alpha=x,z$ is the oscillating direction of the electric field $\boldsymbol E$, and $\frac{1}{\tau}$ denotes the linewidth broadening induced by impurity scattering~\cite{Wang2021, M.Udagawa, P.Hosur, Ashby, Duan2019}.  We choose the Fermi energy at zero energy and set the temperature to be zero, so the index $s=-1$ and $s'=1$.  For the linearly polarized light acting on the system, the optical absorption is proportional to the real part of $\sigma_{\alpha\alpha}$.  

\begin{figure}
	\includegraphics[width=9.2cm]{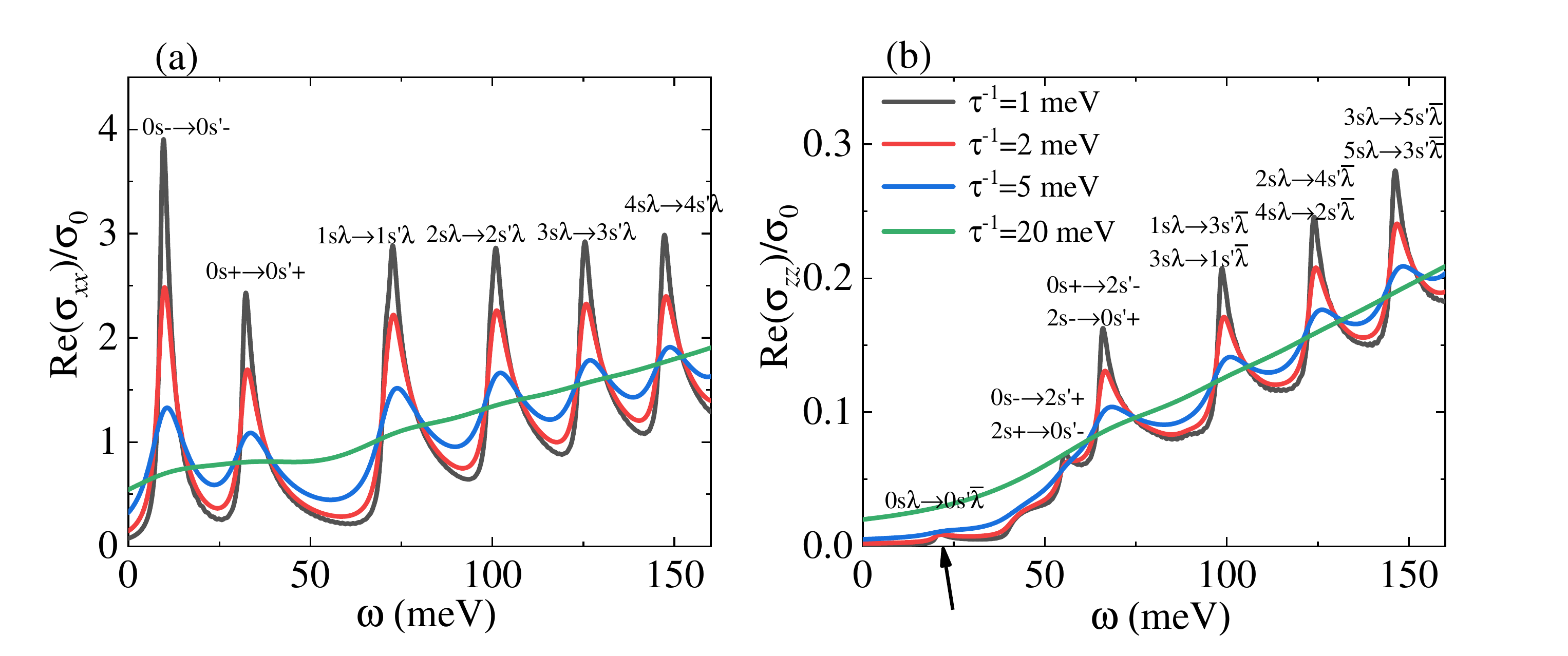}
	\caption{(Color online) The magneto-optical conductivity Re($\sigma_{\alpha\alpha}$) (in units of $\sigma_0=\frac{e^2}{2\pi}$) vs the photon frequency $\omega$ for different $\tau^{-1}$, with $\alpha=x$ in (a) and $\alpha=z$ in (b).  The LL transitions are labeled for each resonant peak, where a weak peak is indicated by the arrow.  The legends are the same in both figures.  We take the same parameters as those in Fig.~\ref{Fig1}(a). }
	\label{Fig2}	
\end{figure}

The numerical results of the magneto-optical conductivity Re$(\sigma_{\alpha\alpha})$ are plotted in Fig.~\ref{Fig2} for different strengths of linewidth broadening.  In the limit of small $\tau^{-1}$ (e.g., 1 meV), one observes a series of resonant peaks sitting on an increasing background~\cite{Ashby}, which is due to the dispersive LLs in the 3D system.  When $\tau^{-1}$ is strong, e.g., $\tau^{-1}=20$ meV, the system enters into the diffusive metallic regime, and the resonant peaks disappear.   
 
The resonant peaks occur at the photon frequency  $\omega=\varepsilon_{n's'\lambda'}-\varepsilon_{ns\lambda}$ at the van Hove singularity $k_x=0$.  The corresponding selection rules can thus be determined for each resonant peak by explicitly matching the energy difference between the initial and final LLs.  
In Fig.~\ref{Fig2}, we carefully label each LL transition, and observe that 
the selection rules are $n\rightarrow n$ in the longitudinal conductivity Re$(\sigma_{xx})$, and $n\rightarrow n\pm2$ in the transverse conductivity Re$(\sigma_{zz})$.
The former selection rules are the same as previous magneto-optic studies of Dirac fermions when $\boldsymbol E\parallel \boldsymbol B$~\cite{L.You, Mukherjee, Wang2020, Duan2020}, whereas the latter ones are distinctively different from the previous studies of Dirac fermions when $\boldsymbol E\perp\boldsymbol B$, with conventional selection rules given as $n\rightarrow n\pm1$~\cite{L.You, Ashby, Duan2019, Mukherjee, Wang2020, Duan2020, R.Y.Chen, Z.G.Chen, Y.Jiang2017, Wang2017}. 
 
It is interesting to note that, due to the broken twofold degeneracy of the zeroth LLs,  the zeroth LL transitions show two different resonant peaks both in Re$(\sigma_{xx})$ and  Re$(\sigma_{zz})$. In addition,  in Re$(\sigma_{zz})$, there exists a weak peak with the LL transition $0s\lambda\rightarrow0s'\bar\lambda$, as indicated by the arrow.  

\begin{figure}
	\includegraphics[width=9.2cm]{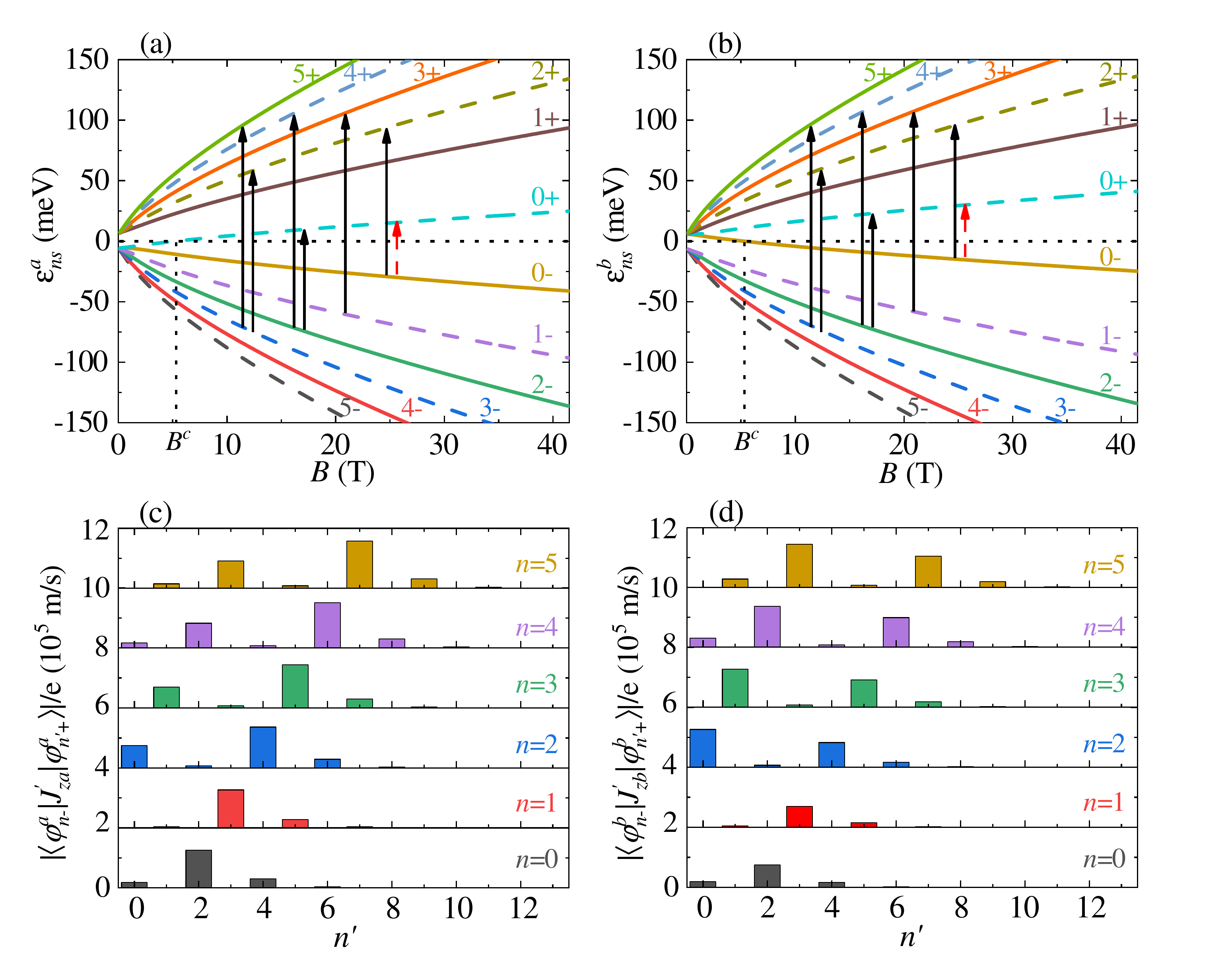} 
	\caption{(Color online) The LLs vs the magnetic field $B$, with $\varepsilon_{ns}^a$ of $H_{aB}'$ in (a) and $\varepsilon_{ns}^b$ of $H_{bB}'$ in (b).  The even-(odd-)parity LLs are labeled as solid (dashed) lines.  The arrows indicate the dominated LL transitions in Re$(\sigma_{zz})$, with the red dashed one related to the weak peak in Fig.~\ref{Fig2}(b).  The matrix element $\langle\varphi_{n-}^{a(b)}|J_{za(b)}'|\varphi_{n'+}^{a(b)}\rangle$ between the different initial states $\varphi_{n-}^{a(b)}$ and final states $\varphi_{n'+}^{a(b)}$ in (c), (d).  The magnetic field is set as $B=10.3$ T. For clarity, the neighboring bars are shifted vertically by $2\times10^5$. }
	\label{Fig3}	
\end{figure}

\subsection{Selection Rules}

To understand the behavior of the magneto-optical conductivity and especially the unconventional selection rules, it would be instructive to consider the Hamiltonian $H(\boldsymbol k)$ at the van Hove singularity $k_x=0$.  At this point, $H(\boldsymbol k)$ can be block diagonalized through a unitary matrix $U=\frac{1}{\sqrt{2}}(\sigma_x + \sigma_z)\otimes \tau_z$:
\begin{align}
H'=UH(k_x=0)U^\dagger=H_a'\oplus H_b', 
\end{align}
where the sub-Hamiltonians $H_a'$ and $H_b'$ are written as
\begin{align}
H_{a,b}'=\mp v_zk_z\tau_x-vk_y\tau_y
+(M-\xi k_y^2-\xi_zk_z^2)\tau_z 
\label{Hab}.
\end{align}
Clearly, $H'$ owns the chiral symmetry as ${\cal C}H'{\cal C}^\dagger=-H$~\cite{Asboth}, with the operator ${\cal C}=\sigma_x\otimes\tau_x$. 
If the anisotropy is absent, $\xi=\xi_z$ and $v=v_z$, $H_a'$ and $H_b'$ represent the 2D Chern insulator models that are expanded to the second-order $k_i^2$ terms; the total Hamiltonian $H'$ is equivalent to the low-energy effective model that describes the 2D surface states in the magnetic TI film~\cite{R.Yu}. 
 
We introduce the ladder operators   
$a=\frac{1}{\sqrt 2}(\eta+\frac{\partial}{\partial\eta})$ and 
$a^\dagger=\frac{1}{\sqrt 2}(\eta-\frac{\partial}{\partial\eta})$, 
with the dimensionless parameter $\eta=l_Bk_z-\frac{y}{l_B}$ and the magnetic length $l_B=\frac{1}{\sqrt{eB}}$, so then $H_a'$ becomes 
\begin{align}
H_{aB}'=
\begin{pmatrix}
P-\tilde{Q} &T a^\dagger-S a
\\
T a-S a^\dagger
&-P+\tilde{Q}
\end{pmatrix}, 
\label{HaB}
\end{align}
with $\tilde{Q}=Q a^\dagger a-\frac{R}{2}(a^2+a^{\dagger 2})$. 
$H_{bB}'$ can be obtained from $H_{aB}'$ by exchanging $S$ and $T$. 
Here the parameters are defined as
$P=M-\frac{\xi+\xi_z}{2l_B^2}$, 
$(Q,R)=\frac{\xi\pm \xi_z}{l_B^2}$, 
$(S,T)=\frac{v \pm v_z}{\sqrt2 l_B}$.  $H_{a/bB}'$ can be diagonalized in the Hilbert space spanned by $|n\rangle$ that is defined as $a^\dagger a|n\rangle =n|n\rangle$~\cite{M.Koshino, T.Devakul}.  In the calculations, one can truncate the Hilbert space at a cutoff $N_c$, which is set as $N_c=200$ (see Sec.IV of SM~\cite{SM}).  

\begin{figure*} 
	\centering
	\includegraphics[width=18cm]{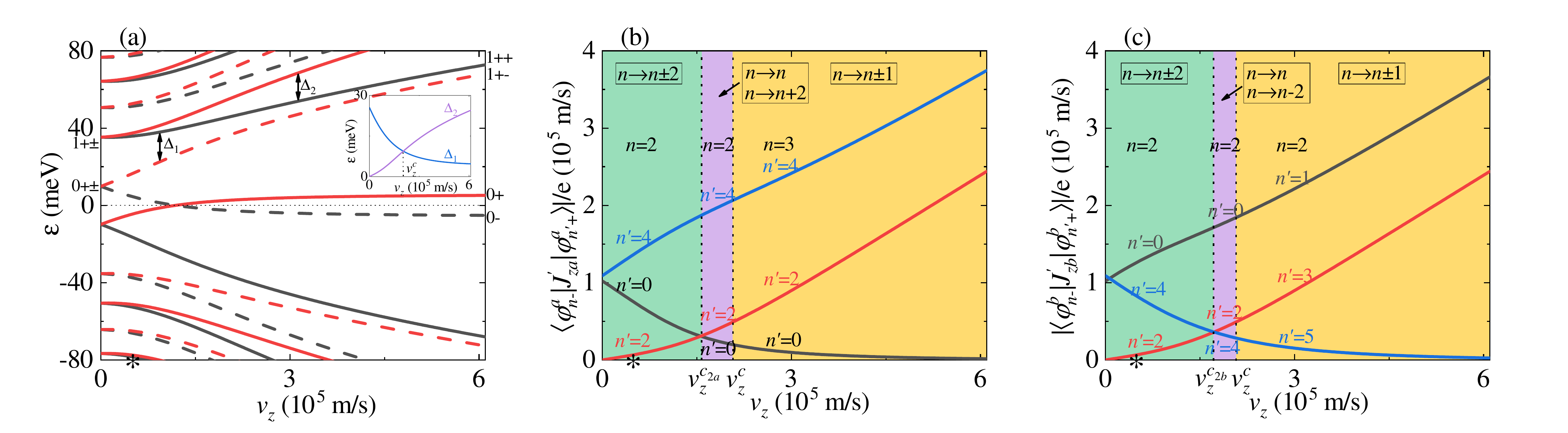}
	\caption{(Color online) The LL flowing with $v_z$ in (a).  The black (red) lines represent $\varepsilon_{ns}^{a(b)}$, and the solid (dashed) lines denote the even-(odd-)parity LLs.  The inset shows the energy differences, $\Delta_1$ and $\Delta_2$.  The matrix element $\langle\varphi_{n-}^{a(b)}|J_{za(b)}'|\varphi_{n'+}^{a(b)}\rangle$ with $v_z$ in (b), (c), where the initial and final state index, $n$ and $n'$, as well as the selection rules are labeled in different regions.  The critical velocity $v_z^{c_1}$ is the same as that in (a), and $v_z^{c_{2a(b)}}$ marks the point where the matrix element inversion occurs.  We set the magnetic field as $B=10.3$ T and mark the experimental value $v_z=0.5\times10^5$ m/s~\cite{Y.Jiang2020} with asterisks. } 
	\label{Fig4}	
\end{figure*} 

The LL energies of $H_{aB}'$ and $H_{bB}'$ are presented as a function of $B$ in Figs.~\ref{Fig3}(a) and~\ref{Fig3}(b), respectively, in which $\varepsilon_{ns}^a$ together with $\varepsilon_{ns}^b$ agree well with the results in Fig.~\ref{Fig1}(b) by a lattice model calculation.  We observe that particle-hole symmetry is broken in $H_{a(b)B}'$, but is still preserved in the total Hamiltonian $H_{aB}'\oplus H_{bB}'$.  This can be seen from the relation $\tau_x H_{aB}'\tau_x=-H_{bB}'$.  Indeed, this relation guarantees that the energies and wavefunctions of the two sub-Hamiltonians are related: $\varepsilon_{ns}^a=-\varepsilon_{n\bar s}^b$ and $\varphi_{ns}^a=\tau_x \varphi_{n\bar s}^b$, with $\bar s=-s$.  Actually, in Fig.~\ref{Fig1}(b), the $(n+-)$ and $(n++)$ LLs with $n>0$ are equivalent to the $n+$ LLs in $H_{aB}'$ and $H_{bB}'$, respectively.  

With the same unitary matrix $U$, the current density operator matrix at $k_x=0$ is transformed as
\begin{align}
&J_x'=UJ_xU^\dagger=-ev\sigma_x\otimes\tau_x, 
\label{J_x}
\\
&J_z'=UJ_zU^\dagger=J_{za}'\oplus J_{zb}', 
\label{J_z}
\end{align}
with 
\begin{align}
J_{za,b}'=\mp ev_z\tau_x-\frac{\sqrt 2e\xi_z}{l_B}(a+a^\dagger)\tau_z. 
\end{align}
As $J_x'$ in Eq.~(\ref{J_x}) is anti-diagonal, its nonzero matrix element must appear between the states of different sub-Hamiltonians.  Since
$\langle \varphi_{n+}^{a(b)}|J_x'|\varphi_{n'-}^{b(a)}\rangle
=-ev\delta_{nn'}$,  the selection rules must be $n\rightarrow n$ and $\lambda'=\lambda$.

For $J_z'$ in Eq.~(\ref{J_z}), since it is block diagonal,  its nonzero matrix element must appear between the states of the same sub-Hamiltonians.  Now each sub-Hamiltonian $H_{a(b)B}'$ owns the parity symmetry with the operator ${\cal P}_a=(-1)^{a^\dagger a}\tau_z$ commuting with $H_{a(b)B}'$~\cite{T.Devakul}. 
Therefore, each LL owns definite even or odd parity, as labeled in Figs.~\ref{Fig3}(a) and~\ref{Fig3}(b) by solid or dashed lines, respectively.  We can see that the neighboring LLs always own opposite parities.  On the other hand, as ${\cal P}_a J_{za(b)}'{\cal P}_a=-J_{za(b)}'$, $J_{za(b)}'$ owns odd parity and its matrix element  $\langle\varphi_{n-}^{a(b)}|J_{za(b)}'|\varphi_{n'+}^{a(b)}\rangle$ is nonvanishing only when the $n-$ and $n'+$ states own opposite parities (see Sec. V of  SM~\cite{SM}).  Thus due to the peculiar distribution of parity carried by the LLs, the nonvanishing matrix element requires $n'-n=2l$, with $l$ being the integer.  This is also demonstrated by the numerical results in Figs.~\ref{Fig3}(c) and~\ref{Fig3}(d).  We see that the matrix elements take large values only when $n'=n\pm2$, and decrease quickly to zero when $n'$ is away from $n$.  Therefore one can determine that the selection rules in Re$(\sigma_{zz})$ are $n\rightarrow n\pm2$ and $\lambda'=-\lambda$.

\subsection{Anisotropy}

As 3D ZrTe$_5$ is highly anisotropy, we investigate its role in forming the LLs and selection rules by changing the Fermi velocity $v_z$.  Experimentally, tuning $v_z$ is quite feasible, because it represents the hopping strength in the $z$ direction and is expected to increase when the external pressure is applied along the $z$ direction of the 3D crystal~\cite{Y.Zhou, J.L.Zhang2017}.  In Fig.~\ref{Fig4}(a), the total LLs are plotted with $v_z$, with the even-(odd-) parity LLs denoted by solid (dashed) lines.  One observes a transition in the distribution of parity in the LLs as increasing $v_z$, which can be understood through the chiral symmetry of the sub-Hamiltonian as follows. 

In the limit $v_z=0$, we have $S=T$ and the sub-Hamiltonian $H_{a(b)B}'$ owns the chiral symmetry $\tau_x H_{a(b)B}'\tau_x=-H_{a(b)B}'$.  The chiral symmetry guarantees that each sub-Hamiltonian has two zeroth LLs but with opposite parities.  
When $v_z$ increases from zero, the chiral symmetry of the sub-Hamiltonian is broken, but the two zeroth LLs still persist, until $v_z$ reaches a critical value. 
After $v_z$ crosses the critical value, one of the two zeroth LLs flows away from its partner, and becomes an $n=1$ LL.  Thus we are left with only one single zeroth LL for each sub-Hamiltonian, leading to the $n+$ and $n-$ LLs carrying the same parities.  If the system is nearly isotropic, $v_z\sim v$, we have $S\gg T,\tilde{Q}$ and the sub-Hamiltonian $H_{a/bB}'$ is analogous to that in graphene~\cite{Z.Jiang, T.Morimoto, M.O.Goerbig} and other Dirac electron systems~\cite{C.J.Tabert}, .

To quantitatively characterize the transition, we define $\Delta_1$ and $\Delta_2$ as the energy differences between the neighboring LLs, as indicated in Fig.~\ref{Fig4}(a).  The inset of Fig.~\ref{Fig4}(a) presents the relative magnitude of $\Delta_1$ and $\Delta_2$, in which $\Delta_1=\Delta_2$ occurs at $v_z=v_z^c$.  This critical $v_z^c$ characterizes when the two zeroth LLs change to be one zeroth LL.  The transition also exists in higher $n>0$ LLs (see Sec. VII of SM~\cite{SM}).  Explicitly, when $v_z$ increases, the $ns+$ LL flows to the $(n+1,s-)$ LL, with a direct consequence being that the index of the two LLs should be relabeled as $(n+1,s-)$ and $(n+1,s+)$, respectively.  

Next, we turn to the selection rules in Re$(\sigma_{zz})$.  In Fig.~\ref{Fig4}(b), for the matrix element $\langle\varphi_{n-}^a|J_{za}'|\varphi_{n'+}^a\rangle$, when $v_z\ll v$, the initial and final state index are chosen as $n=2$ and $n'=0,2,4$, respectively.  Clearly the matrix element with $n'=0,4$ takes a relatively large value, indicating the unconventional selection rules $n\rightarrow n\pm2$.  When $v_z$ increases to be larger than $v_z^{c_{2a}}$, the matrix element with $n'=2$ overwhelms that with $n'=0$, thus the selection rules become $n\rightarrow n$ and $n\rightarrow n+2$.  Further increasing $v_z$ to cross $v_z^c$, the LLs regroup and the initial state index will be increased by one but the final state index remains unchanged.  Consequently, the parity of the $n-$ LL changes, while the parity of the $n'+$ LL remains unchanged, as seen in Fig.~\ref{Fig4}(a).  Therefore, the conventional selection rules $n\rightarrow n\pm1$ are recovered.  A similar change of the selection rules can also be found for the matrix element  $\langle\varphi_{n-}^b|J_{zb}'|\varphi_{n'+}^b\rangle$ in Fig.~\ref{Fig4}(c) and those with a higher index (see Sec. VIII of SM~\cite{SM}).

\subsection{Zeroth LLs versus the magnetic field}

As analyzed above, zeroth LLs play an important role in forming the unconventional selection rules in ZrTe$_5$.  Here, we make a detailed study of the anomalous behavior of the zeroth LLs with the magnetic field $B$.  In Figs.~\ref{Fig3}(a) and ~\ref{Fig3}(b), when $B$ increases, the $0-$ ($0+$) LLs in $H_{a(b)B}'$ changes its sign, and crosses the zero energy at the same critical magnetic field $B^c=5.24$ T.  This can be understood from the effective Dirac mass term $P$ of $H_{aB}'$ in Eq.~(\ref{HaB}).  The ``mass'' $P$ consists of the original Dirac mass $M$ and a part proportional to an external magnetic field.  For the case of $M>0$, the effective mass $P$ changes from positive to negative with increasing magnetic field, leading to the sign inversion of the lowest LL.  However, for the case of $M<0$, there would be no sign inversion of the lowest LLs.  Therefore, the intercept of the lowest LLs with the zero energy is closely connected to the bulk band inversion, which gives an important signature of a strong TI phase in ZrTe$_5$.  In a recent thermoelectric effect~\cite{J.L.Zhang2019} and another magnetoinfrared spectroscopy study~\cite{Z.G.Chen}, the intercept of the lowest LLs in ZrTe$_5$ with zero energy was also demonstrated, but under a perpendicular magnetic field.  The physical mechanisms were both attributed to the strong Zeeman splitting, with the critical magnetic field estimated to be around 13 T~\cite{J.L.Zhang2019} and 17 T~\cite{Z.G.Chen}, much higher than the present study.  More discussions regarding the zeroth LLs are presented in Secs. II and III of SM~\cite{SM}.

\section{Discussions and Conclusions}

In this paper, by performing full quantum mechanical calculations, we study the magneto-optics in 3D ZrTe$_5$ under an in-plane magnetic field and reveal the unconventional selection rules $n\rightarrow n\pm2$ in Re$(\sigma_{zz})$.  It should be mentioned that, several studies have been performed for 3D Weyl semimetals under an in-plane magnetic field,  and it was found that the chiral anomaly can engender the planar Hall effect~\cite{A.A.Burkov, S.Nandy, S.Ghosh}.   However, these studies were mainly in the semiclassical framework, and thus quantum effects are not fully uncovered.
We also mention that, in the optical absorption spectra of MoS$_2$~\cite{F.Rose} and the photocurrent of bilayer graphene~\cite{L.Ju}, similar unconventional selection rules were reported, which were attributed to a quite different mechanism, the high-order trigonal warping effect.  Moreover, in their studies~\cite{F.Rose, L.Ju}, the unconventional selection rules were only connected to very weak peaks and thus are difficult to be observed in experiments.

{\color{red} For the electron-electron correlations, it is estimated that within the mean-field framework~\cite{F.Qin}, when the magnetic field is above a critical value, the correlations can open up a gap with a magnitude of meV around the Fermi level.  However, the parity of the LLs will not be changed by such a gap opening, and thus the selection rules would not be affected.  We also demonstrate that the selection rules show a certain robustness to the weak in-plane Zeeman splittings (see Sec.~IX of SM~\cite{SM}).} 

We make some comparisons with a recent experiment in ZrTe$_5$~\cite{Y.Jiang2020} where the magnetic field was applied along the $a$ axis: (i) The transition energy extracted from the magnetoinfrared spectra clearly exhibits two branches in the zeroth LL transitions and one branch in the $n\geq1$ LL transitions, which are consistent with our theoretical results; (ii) the magnetic field implemented in the experiment was in the range $6-17.5$ T with the asymptotic exponent for the LLs being $\frac{1}{2}$, whereas our calculations indicate that the asymptotic exponents for the LLs are quite different, and may require stronger magnetic fields to test the discrepancy.  Therefore, further experimental works in ZrTe$_5$ are expected.  We also hope that our results can be extended to other layered and weak-coupling 3D topological materials, such as HfTe$_5$~\cite{Weng, P.Wang, Galeski}, Bi$_2$Te$_3$, and Sb$_2$Te$_3$~\cite{H.Zhang, C.X.Liu}.

\section{Acknowledgment} 

This work was supported by the National Key Research and Development Program of Ministry of Science and Technology (No. 2021YFA1200700) and the Natural Science Foundation of China (Grants No. 11804122, 11905054, and 12275075), the China Postdoctoral Science Foundation (Grant No. 2021M690970), and the Fundamental Research Funds for the Central Universities of China.

\end{document}